\documentclass[prd,aps,tightenlines,a4paper,showkeys,showpacs,12pt]{revtex4}

\usepackage{graphicx}
\usepackage{bm}

\usepackage{url}
\usepackage{amsmath}
\usepackage{multirow}

\usepackage[export]{adjustbox}

\newcommand{\muas}[0]{\hbox{\rm $\mu$as}}

\def\ve#1{{\bf #1}}

\def\aap{{Astron.~Astrophys.}}

\begin{document}

\title{Gaia-like astrometry and gravitational waves}

\author{Sergei A. \surname{Klioner}}

\affiliation{Lohrmann-Observatorium, Technische Universit\"at Dresden,
01062 Dresden, Germany}

\begin{abstract}
This paper discusses the effects of gravitational waves on
high-accuracy astrometric observations such as those delivered by
Gaia.  Depending on the frequency of gravitational waves, two regimes
for the influence of gravitational waves on astrometric data are
identified: the regime when the effects of gravitational waves
directly influence the derived proper motions of astrometric sources
and the regime when those effects mostly appear in the residuals of
the standard astrometric solution. The paper is focused on the
second regime while the known results for the first regime are briefly
summarized.

The deflection of light due to a plane gravitational wave is then
discussed.  Starting from a model for the deflection we derive the
corresponding partial derivatives and summarize some ideas for the
search strategy of such signals in high-accuracy astrometric data. In
order to reduce the dimensionality of the parameter space the use of
vector spherical harmonics is suggested and explained. The explicit
formulas for the VSH expansion of the astrometric signal of a plain
gravitational wave are derived.

Finally, potential sensitivity of Gaia astrometric data is
discussed. Potential astrophysical sources of gravitational waves
that can be interesting for astrometric detection are identified.
\end{abstract}


\maketitle

\section{Introduction}
\label{section-Introduction}

The main goal of Gaia-like global astrometry is to determine positions, proper motions
(together with special solutions for non-single stars), and parallaxes of celestial
objects. However, observational astrometric data can be used
to search for other effects. Examples here are various tests of
general relativity (e.g., the PPN parameter $\gamma$ or the quadrupole
light deflection due to Jupiter). One more possibility is to search
for the astrometric signatures of gravitational waves of various
frequencies.

It is known that the prospects for an astrometric detection of
gravitational waves are not very promising (see \citep{Schutz2010} and
references therein). Nevertheless, it is interesting to elaborate the
details of what can be expected from the current microarcsecond
astrometric projects (first of all, from Gaia). It is clear that this
theory, ideas and algorithms will be even more important for the
next-generation sub-microarcsecond astrometric missions that are being
actively discussed
(e.g. \citep{MalbetEtAl2012,2017arXiv170701348T,2016arXiv160907325H})
and hopefully will be realized in the future.  Although any sort of
astrometry can be used to detect gravitational waves, in this paper we
pay special attention to Gaia-like global astrometry from space.

The use of high-accuracy astrometry to detect various kinds of
gravitational waves was discussed by many authors \citep[][and references
 therein]{BraginskyEtAl1990,PyneEtAl1996,GwinnEtAl1997,2010IAUS..261..345M,BookFlanagan2011,MignardKlioner2012}.
The first subject of these studies was the effect of stochastic
background of ultra-low-frequency primordial gravitational waves
on the propagation of light as seen in astrometry,
pulsar timing and other observational techniques. Then, triggered by some
false claims in the literature, astrometric effects of gravitational
waves from localized sources have been studied in great details
\citep[][and references
  therein]{1998PhRvD..58d4003D,1999PhRvD..59h4023K,2001LNP...562..141B,Schutz2010}.
Starting from the very beginning of the space astrometry project Gaia
it was clear that Gaia can potentially put an interesting limit on the
energy flux of primordial gravitational waves
\citep{ESA2000,Klioner2007,MignardKlioner2012}. Finally, the idea to
search for higher-frequency gravitational waves in the residuals of
the astrometric solution of Gaia emerged around 2008
\citep{MignardKlioner2010,Klioner2013,Klioner2015}. An early attempt to
implement an algorithm to search for gravitational waves in astrometric data
was undertaken in \citep{Geyer2014}.

Gaia implements a special sort of astrometric instrument: a scanning
astrometric space telescope \citep{ESA2000,PrustiEtAl2016}.
The observational data represent exact times of observations of
astrometric sources at some predefined fiducial lines in two fields of
view. From these data several kinds of parameters should be estimated:
source parameters (the standard parametrization includes 5 parameters
per source: two components of the position, two components of the
proper motion, and parallax), attitude parameters defining the
attitude of the instrument as function of time, and calibration
parameters describing properties of the observational instrument (if
needed, as function of time as well). All these parameters should be determined
in a complicated robust least squares estimation
process.  The estimated values of the parameters are called below
``astrometric solution'' or ``standard astrometric solution''.
In principle, the values of parameters don't depend (or depend
only insignificantly) on the details of the estimation process.
In Gaia, a special sort of the estimation process called
``Astrometric Global Iterative Solution'' (AGIS) \citep{LindegrenEtAl2012}
is used. AGIS is flexible and computationally extremely efficient.
It is the invention of AGIS that made Gaia computationally feasible.
However, on the matter of principle, one can think of other approaches.

In the case of Gaia, additional information can be
extracted from observational data by including additional effects
directly into the astrometric model and fitting the corresponding
parameters directly in AGIS. However, this is not always the
best way, especially if the corresponding additional effect is
substantially non-linear with respect to the parameters that should be
determined from observations. This is the case for the astrometric
effects caused by gravitational waves, where a non-linear global
optimization is required. The AGIS method is suitable for non-linear
local optimization \citep{LindegrenEtAl2012}.  It means that AGIS
assumes that the initial values of all parameters are close enough to
the true ones. In the case of gravitational waves, no reasonable
apriori values for the wave parameters are known.  For this reason, a
different strategy is needed.  It is the purpose of this paper to
discuss how high-accuracy astrometric data can be used for the search
for gravitational waves, to give some theoretical background of the
astrometric signals from gravitational waves, and to summarize some
ideas and algorithms useful for such a search. This is the first
paper in a series of publications devoted to the search of
gravitational waves in high-accuracy astrometric data. Subsequent
papers will give further details of the algorithms and their
implementations, describe detailed analysis of the interaction of the
gravitational-wave signal and the standard astrometric solution, and,
finally, show the results of the search attempts using
the real astrometric data of Gaia.

Section~\ref{section-two-regimes} sketches two regimes of the
interaction of a gravitational wave and the astrometric solution.
Section~\ref{section-deflection} is devoted to the deflection of light
due to a plane gravitational wave of a given frequency and direction.
The formulas for the partial derivatives with respect to the
parameters of the gravitational wave are given in
Section~\ref{section-partial-derivatives}. A sketch of the algorithm that uses
the vector spherical harmonics (VSH) to reduce the parameter space
for the global optimization problem is presented in
Sections~\ref{section-vhs} and \ref{sect-VSH-expansion}. Section~\ref{section-frequency-limits}
contains a discussion for the frequency range accessible by Gaia
astrometry as well as a basic sensitivity analysis for Gaia.  In
Section~\ref{section-sources}, it is argued that binary supermassive
black holes in remote galaxies seem to be the most promising sources for
astrometric detection of gravitational waves. Finally,
Section~\ref{section-conclusion} summarizes the results and discusses the
prospects of the field.

\section{Regimes of the interaction between gravitational waves and the astrometric solution}
\label{section-two-regimes}

It is generally clear that gravitational waves being time-dependent
(periodic) gravitational fields cause a time-dependent (periodic)
deflection of light. This additional deflection of light leads to
time-dependent (periodic) shifts of apparent positions of celestial
sources. This is confirmed by detailed analytical studies of the
effect \citep[][and references
  therein]{PyneEtAl1996,BookFlanagan2011}.  As long as gravitational
waves remain undetected and their characteristics are unknown, it is
computationally impossible to include the effects of gravitational
waves in the corresponding astrometric models. The reason for this is
the substantially non-linear character of the model for the astrometric effects of
gravitational waves. Therefore, the effects of gravitational waves
remain unmodeled and may influence astrometric solutions. Depending on
the relation between the period of a gravitational wave and the time
span covered by observations one can apriori see two ways how the
effects of a gravitational wave can influence the astrometric
solution.

If the period of a gravitational wave is much larger than the time
span covered by the observational data, the time-dependent deflection
of light can be sufficiently well described by linear motions of
sources on the sky. A linear motion is a part of the standard
astrometric model for the apparent motion of celestial sources. In
this case a significant part of the effects of gravitational waves
changes the proper motions of the sources and only a small part of the
effects (if at all) goes to the residuals of an astrometric
solution. The theory of these changes due to gravitational waves has
been developed by a number of authors
\citep{BraginskyEtAl1990,PyneEtAl1996,BookFlanagan2011}. The theory
was used to give upper estimates for the energy flux of the primordial
gravitational waves using geodetic VLBI observations
\citep{GwinnEtAl1997,TitovLambertGontier2010,TitovLambert2013,MignardKlioner2012}. An
estimate of what can be expected from Gaia in this area was given in
\citep{MignardKlioner2012} for the expected pre-launch accuracy and was
corrected for the post-launch estimation of the accuracy in
\citep{Klioner2015}: Gaia, for the nominal observation period of 5 years,
is expected to give the upper estimate of
the energy of the gravitational waves at the level of $\Omega_{\rm
  gw}<0.00012\,h_{100}^{-2}$ of the closure energy of the Universe for
frequencies $\nu<6.4\times10^{-9}$~Hz.  Here $h_{100}=H/(100\, {\rm km}/{\rm
  s}\,{\rm Mpc}^{-1})$, $H$ being the Hubble constant. This is at
least 80 times better than the current best estimate provided by
geodetic VLBI \citep[][see the discussion of the VLBI results in Section~7.2]{MignardKlioner2012}.
This regime is not further discussed in this paper.

If the period of a gravitational wave is considerably smaller than the
time span of the data, an important part of its signal should go to
the residuals of the astrometric solution. As discussed e.g. in
\citep{Klioner2014,2017A&A...603A..45B}, if one neglects the
second-order effects due to the finite size of the fields of view the
across-scan part of an {\it arbitrary}\/ signal only modifies the
across-scan attitude of the solution. Similarly, half of the sum of
the along-scan signal in two fields of view also only modifies the
along-scan attitude.  These effects on the attitude as well as
``differential'' effects due to the finite size of the fields of view
do influence the astrometric parameters, but only as second-order
effects.  It is only the remaining along-scan signal (the deviation of the
signal in each field of view from the mean value between the two fields of
view) that directly influences the effective basic angle between the
fields of view and, therefore, potentially alters the astrometric
solution and its along-scan residuals. Source parameters, attitude and
standard calibration parameters cannot {\it fully}\/ absorb a periodic
signal. Most of the gravitational wave signal is expected to survive
in the residuals. This represents the second regime of the interaction
of a gravitational wave signal and astrometric solution.

In principle, one can think of a third regime when the period of
gravitational waves is comparable with the time span covered with
observations.  In this case both the source parameters
and the residuals should be affected by the gravitational wave
signal. One could expect that this is the most difficult case.

Obviously, a detailed investigation is needed to clarify the
quantitative characteristics of these regimes. By means of dedicated
numerical simulations of the astrometric solution in the presence of
gravitational wave signal in the data one can analyze the exact character
of the interaction between the gravitational wave and the astrometric
solution. The results of these simulations will be published
separately.

\section{The deflection formula from a plane gravitational wave}
\label{section-deflection}

Here we consider an observer moving in the solar system and a plane
gravitational wave propagating through the solar system and disturbing
the metric tensor in the vicinity of the solar system. The presence of a
gravitational wave results in an additional deflection of light from
remote sources (stars and quasars). In the case of a single plane
gravitational wave of a fixed frequency, the additional deflection of
light is periodic in time, follows a certain pattern with respect to
the observed direction and can be directly detected by the observer.

The deflection of a light ray due to a plane gravitational wave is
discussed in details be many authors \citep[][and reference
  therein]{PyneEtAl1996,BookFlanagan2011,Schutz2009}.  It is well
known that the deflection of light due to a gravitational wave depends
on the strain of the gravitational wave both at the observer and at
the source of light. In case of space astrometry one observes stars in
our Galaxy and compact extragalactic objects (bright stars in the nearby
galaxies, compact remote galaxies, and QSOs), so that the
gravitational wave in question does have some non-zero strain at the
location of a typical source. Those source-related effects cannot be
directly computed since the distances to the stars are not known
sufficiently accurate (even using the results of astrometry). On the
other hand, because observations are performed according to a certain
observational schedule \citep{PrustiEtAl2016} unrelated to the
gravitational wave and because the stars seen close to each other by
an observer have in most cases different distances, the source-related
effects only represent some additional random noise.  In the case of
Gaia, this additional random noise can be expected to be several orders of magnitude
lower than the normal random observational noise (e.g. about 300 \muas\ for
a star of Gaia magnitude $G=15$ \citep{2016A&A...595A...4L,2016A&A...595A...3F}).
From this point of view, astrometric detection of gravitational waves is independent
of the distant source limit discussed e.g. in Section II.G of \citep{BookFlanagan2011}.
The source-related effects will be completely ignored in this paper.
If a better approach for the source-related effects in astrometry can be
found is an open question to be further investigated.

Standard relativistic astrometric models \citep[e.g.,][]{Klioner2003}
take into account the deflection of light due to solar system bodies.
In the spirit of the post-Newtonian approximation scheme used in those
models, the perturbations of the metric tensor due to a gravitational
wave can be considered as purely additive. This also agrees with the fact that the
gravitational fields due to gravitational waves are weak and can be considered in
a linear regime. The effects that are
neglected in this linear approximation are utterly small and can be safely
neglected even at the accuracy level of 0.001 \muas\ for any realistic
gravitational waves. The level of 1 nanoarcsecond (0.001 \muas) is
mentioned here as an ultimate accuracy goal of all future astrometric
projects discussed up to now.

Both \citet{PyneEtAl1996} and \citet{BookFlanagan2011} considered the
case of an observer at rest, e.g. at the barycentre of solar system,
so that the barycentric coordinates of the observer $\ve{x}_{\rm
  obs}$ vanish. On the contrary, for the case of a plane gravitational wave
observed over longer period of time it is important to take into
account the phase changes of the observed wave due to the barycentric
motion of observer in the solar system. Here we consider the practical case of slowly moving
observer, so that its barycentric velocity is much smaller that the velocity of light $c$.

Using all these considerations the variation of the direction towards an astrometric 
source due to a plane gravitational wave can be written as \citep{PyneEtAl1996,BookFlanagan2011}
\begin{eqnarray}
\label{delta-u}
\delta u^i&=&{u^i+p^i\over 2\,(1+\ve{u}\cdot\ve{p})}\,h_{jk}\,u^j\,u^k
-{1\over 2}\,h_{ij}\,u^j\,,
\end{eqnarray}
\noindent
where $h_{ij}$ is the metric perturbation due to the gravitational wave
\citep{Schutz2009,PyneEtAl1996}
\begin{eqnarray}
\label{hij}
h_{ij}&=&h^+\,p_{ij}^+\,\cos\bigr(2\pi\,\nu\,(t-{1\over c}\,\ve{p}\cdot\ve{x}_{\rm obs}-t^+)\bigr)
+h^\times\,p_{ij}^\times\,\cos\bigr(2\pi\,\nu\,(t-{1\over c}\,\ve{p}\cdot\ve{x}_{\rm obs}-t^\times)\bigr)\,,
\\
\label{hij-+}
p_{ij}^+&=&\left({\bf P}\,{\bf e}^+\,{\bf P}^{\rm T}\right)_{ij}\,,
\\
\label{hij-times}
p_{ij}^\times&=&\left({\bf P}\,{\bf e}^\times\,{\bf P}^{\rm T}\right)_{ij}\,,
\end{eqnarray}
\noindent
$\ve{p}$ is the direction of propagation of the gravitational wave,
${\bf e}^+$ and ${\bf e}^\times$ are the polarization matrices, $h^+$
and $h^\times$ are the corresponding strain parameters, ${\bf P}$ is a
special rotational matrix, and $\nu$ is the frequency of the
gravitational wave.  The notations are further explained below.

Vector $\ve{u}$ is the direction from the observer to the source
at the moment of observation.  In our approximation we can
neglect the light deflection due to solar system and consider that
$\ve{u}=-\ve{k}$, where $\ve{k}$ is the coordinate vector from the
source to the observer at the moment of observation $t$ as calculated
by the relativistic astrometric model (e.g. Gaia Relativity Model
(GREM) \citep{Klioner2003,2004PhRvD..69l4001K}) from the source
parameters and the position of the observer $\ve{x}_{\rm
  obs}=\ve{x}_{\rm obs}(t)$.  Both source parameters and $\ve{x}_{\rm
  obs}$ are defined in the underlying relativistic reference system
called Barycentric Celestial Reference System (BCRS) and described
e.g. in \citep{SoffelEtAl2003}.  The correction $\delta\ve{u}$ is
perpendicular to $\ve{u}$ ($\delta\ve{u}\cdot\ve{u}=0$) and represent
the perturbation of the direction towards the source as observed by a
fictitious observer that is at rest with respect to the BCRS and
co-located with the real (moving) observer.  From the point of view of
the relativistic model used for Gaia, $\delta u^i$ given by
Eq.~(\ref{delta-u}) should be added to the direction of light
propagation before correcting for aberration \citep[Section~5]{Klioner2003}.

A plane gravitational wave is fully defined by 7 scalar parameters: $\nu$ is
the frequency of the gravitational wave, $h^+$ and $h^\times$ are the
amplitudes of the two polarization modes of the gravitational wave,
$t^+$ and $t^\times$ are the time epochs defining the phases of the
two polarization modes, and $\ve{p}$ is the direction of propagation of
the gravitational wave that can be parametrized as 
\begin{eqnarray}
\label{p-alpha-delta}
\ve{p}&=&
\begin{pmatrix}
\cos\alpha_{\rm gw}\,\cos\delta_{\rm gw}\\
\sin\alpha_{\rm gw}\,\cos\delta_{\rm gw}\\
\sin\delta_{\rm gw}
\end{pmatrix}
\,,
\end{eqnarray}
\noindent
where $(\alpha_{\rm gw},\delta_{\rm gw})$ are the right ascension and declination of
the direction of propagation.
It is clear that the right ascension and declination of the source of the gravitational wave read
\begin{eqnarray}
&&\alpha_{\rm gw\,source}=\left(\alpha_{\rm gw}+\pi\right)\bmod2\pi\,,
\\
&&\delta_{\rm gw\,source}=-\delta_{\rm gw}\,.
\end{eqnarray}
\noindent
The polarization matrices are defined as:
\begin{eqnarray}
{\bf e}^+&=&
\left(\,\,
\begin{array}{rrr}
1&
\phantom{--}
0&
\phantom{-}
0\\[1pt]
0&-1&0\\[1pt]
0&0&0
\end{array}\,\,
\right),
\\
{\bf e}^\times&=&
\left(\,\,
\begin{array}{rrr}
0&
\phantom{--}
1&
\phantom{-}
0\\[1pt]
1&0&0\\[1pt]
0&0&0
\end{array}\,\,
\right)\,,
\end{eqnarray}
and the rotational matrix ${\bf P}$ is defined as:
\begin{eqnarray}
\label{matrix-P}
{\bf P}&=&{\bf R}_z\left({\pi\over 2}-\alpha_{\rm gw}\right)\,{\bf R}_x\left({\pi\over 2}-\delta_{\rm gw}\right)\,
{\bf R}_z\left(\pi\right)
\noindent\\[5pt]
&=&
\left(\hspace*{-4pt}
{\arraycolsep=7pt
\begin{array}{rrr}
-\sin\alpha_{\rm gw} & -\cos\alpha_{\rm gw}\,\sin\delta_{\rm gw} & \cos\alpha_{\rm gw}\,\cos\delta_{\rm gw}\\
\cos\alpha_{\rm gw} & -\sin\alpha_{\rm gw}\,\sin\delta_{\rm gw} & \sin\alpha_{\rm gw}\,\cos\delta_{\rm gw}\\
0 & \cos\delta_{\rm gw} & \sin\delta_{\rm gw}
\end{array}
}
\hspace*{-4pt}
\right)
\,,
\end{eqnarray}
\noindent
where 
\begin{eqnarray}
\label{rotation-z}
{\bf R}_z(\varepsilon)&=&
\begin{pmatrix}
\cos\varepsilon& \sin\varepsilon&0\\
-\sin\varepsilon&\cos\varepsilon&0\\
0&0&1
\end{pmatrix}\,,
\\
\label{rotation-x}
{\bf R}_x(\varepsilon)&=&
\begin{pmatrix}
1&0&0\cr
0&\cos\varepsilon&\sin\varepsilon\\
0&-\sin\varepsilon&\cos\varepsilon\\
\end{pmatrix}\,.
\end{eqnarray}

Note that ${\bf P}$ is a rotational matrix between the reference system in which
the gravitational wave propagates in the direction of $z$ axis and our normal reference system
in which the propagation direction is $\ve{p}$:
\begin{eqnarray}
\label{p-alpha-delta-via-P}
\ve{p}&=&{\bf P}\,
\begin{pmatrix}
0\\
0\\
1
\end{pmatrix}
\,.
\end{eqnarray}
\noindent
The rightmost rotation ${\bf R}_z\left(\pi\right)$ in the definition
(\ref{matrix-P}) of ${\bf P}$ doesn't change
the matrices $p_{ij}^+$ and $p_{ij}^\times$ as appear in (\ref{hij}).
Therefore, the definition of ${\bf P}$ could be simplified and this
was used e.g. in \citep{PyneEtAl1996}. However, we prefer to retain the definition
(\ref{matrix-P}) since the columns of matrix ${\bf P}$ coincide with the vectors of the
local triad defined by $(\alpha_{\rm gw},\delta_{\rm gw})$
(see e.g. Section~2.1 of \citep{MignardKlioner2012}).

It is convenient to replace the two phases $t^+$ and $t^\times$ by two additional strain parameters:
\begin{eqnarray}
\label{hij-four-amplitudes}
h_{ij}&=&p_{ij}^+\,
\left(h^+_c\cos\Phi+h^+_s\sin\Phi\right)
+p_{ij}^\times\,
\left(h^\times_c\cos\Phi+h^\times_s\sin\Phi\right)
\,,
\\
\label{Phi}
\Phi&=&2\pi\,\nu\,(t-{1\over c}\,\ve{p}\cdot\ve{x}_{\rm obs})\,.
\end{eqnarray}
\noindent
Here both the amplitude and phase are parametrized by four independent parameters
$h^+_c$, $h^+_s$, $h^\times_c$ and $h^\times_s$. Obviously, these parameters depend also on the chosen 
zero-point for the time coordinate $t$.

\section{Partial derivatives and the intrinsic non-linearity of the model}
\label{section-partial-derivatives}

For practical calculations both the correction $\delta u^i$ and its partial derivatives with respect to
the parameters are needed.
The model of the deflection $\delta u^i$ due to a gravitational wave contains a total of 7 parameters.
Four parameters -- the amplitudes $h^+_c$, $h^+_s$, $h^\times_c$ and $h^\times_s$ --
enter the model in a perfectly linear
way, and this can be used to optimize the calculations:
\begin{eqnarray}
\delta u^i&=&
 {\partial \delta u^i\over \partial h^+_c}\,h^+_c
+{\partial \delta u^i\over \partial h^+_s}\,h^+_s
+{\partial \delta u^i\over \partial h^\times_c}\,h^\times_c
+{\partial \delta u^i\over \partial h^\times_s}\,h^\times_s\,,
\end{eqnarray}
\noindent 
where
\begin{eqnarray}
{\partial \delta u^i\over \partial h^+_c}
&=&\delta^i_+\,\cos\Phi
\,,\\
{\partial \delta u^i\over \partial h^+_s}
&=&\delta^i_+\,\sin\Phi
\,,\\
{\partial \delta u^i\over \partial h^\times_c}
&=&\delta^i_\times\,\cos\Phi
\,,\\
{\partial \delta u^i\over \partial h^\times_s}
&=&\delta^i_\times\,\sin\Phi
\,,\\
\label{delta-i-+}
\delta^i_+&=&f^{ijk}\,p_{jk}^+
\,,\\
\label{delta-i-times}
\delta^i_\times&=&f^{ijk}\,p_{jk}^\times
\,,\\
\label{fijk}
f^{ijk}&=&{1\over 2}\,\left({u^i+p^i\over 1+\ve{u}\cdot\ve{p}}\,u^j\,u^k-\delta^{ij}\,u^k\right)
\,.
\end{eqnarray}
\noindent
Note that in the limit $\ve{u}\to-\ve{p}$ there is no degeneracy and
all partial derivatives and $\delta u^i$ go to zero.
The parameters $\alpha_{\rm gw}$ and $\delta_{\rm gw}$ describing the
direction of the gravitational wave 
as well as the frequency $\nu$ of the gravitational wave
enter the model in a substantially
non-linear way. The derivative with respect to $\nu$ is easy to compute as:
\begin{eqnarray}
\label{partial-ui-partialOmega}
{\partial \delta u^i\over \partial\nu}
&=&
{\Phi\over \nu}\,
\left(
-{\partial \delta u^i\over \partial h^+_s}\,h^+_c
+{\partial \delta u^i\over \partial h^+_c}\,h^+_s
-{\partial \delta u^i\over \partial h^\times_s}\,h^\times_c
+{\partial \delta u^i\over \partial h^\times_c}\,h^\times_s
\right)
\,.
\end{eqnarray}
\noindent
The derivatives with respect to $\alpha_{\rm gw}$ and $\delta_{\rm
  gw}$ are straightforward to calculate.  These derivatives cannot
save any calculations of other derivatives and $\delta u^i$ itself and
are not given here explicitly. In astrometry, one usually uses the
differential in right ascension as a true arc
\citep[e.g.][Section~5.1.3]{LindegrenEtAl2012}, thus computing ${1\over \cos\delta_{\rm
    gw}}\,{\partial u^i\over\partial\alpha_{\rm gw}}$ often denoted as
${\partial u^i\over\partial\alpha^*_{\rm gw}}$.  This derivative is
degenerate at the poles $\delta_{\rm gw}=\pm\pi/2$. The degeneracy is
only a problem of parametrization and has no deeper mathematical or
physical meaning. Therefore, for the
fit of $(\alpha_{\rm gw},\delta_{\rm gw})$, one should avoid starting points located too close to the
poles. If this is necessary or happens in the process of iterations of
a non-linear least squares optimizer, one can use e.g. the Scaled
MOdeling of Kinematics (SMOK) as described in \citep[][Appendix
  A]{2014A&A...571A..85M}.

Because of the intrinsic non-linearity of the model and the fact that
apriori we do not have any good initial approximation for all 7
parameters, it is clear that finding optimal values of these
parameters represents a global non-linear optimization problem. Even
if the parameter space has moderate dimensionality, the optimization
problem appears to be computationally difficult.

\section{The use of the vector spherical harmonics to detect gravitational waves in astrometric data}
\label{section-vhs}

To improve the computational complexity of the search for
gravitational waves in astrometric data one should attempt to reduce the
number of non-linear parameters of the model exposed above as much as
possible. One promising way to do this is to use the technique of
vector spherical harmonics (VSH) in combination with an iso-latitude
sky pixelization scheme (e.g. HEALPix \citep{GorskyEtAl2005}) that
allows one to accelerate the computation of VSH fits for given data.

General idea is to use the expansion of a vector field $\delta u^i$ in
vector spherical harmonics (e.g. \citep{MignardKlioner2012}) at any
given moment of time. Such an expansion would allow to detect the
signal from gravitational waves in a rotationally invariant way, so
that both the amplitudes $h^+_c$, $h^+_s$, $h^\times_c$, and
$h^\times_s$ and the direction of gravitational wave $(\alpha_{\rm gw},\delta_{\rm gw})$
are all estimated simultaneously. Clearly, the
VSHs are more suitable to describe {\it time-independent}\/ vector fields of a sphere.
Otherwise VSH coefficients themselves become
time-dependent and one should estimate a function of time for each VSH
coefficient instead of a constant. This is obviously possible, but
would lead to a substantial loss of accuracy.
Fortunately, there is a way to use the
VSH expansion for the vector field $\delta u^i$ in an efficient way
using a simple approximation.

Let us first represent the field $\delta u^i$ in the following way:
\begin{eqnarray}
\label{delta-u-split}
\delta u^i&=&V^i_c\,\cos\Phi+V^i_s\,\sin\Phi\,,
\end{eqnarray} 
\noindent
where $\ve{V}_c$ and $\ve{V}_s$ are two time-independent vector fields
depending on the 6 parameters of the gravitational wave $h^+_c$,
$h^+_s$, $h^\times_c$, $h^\times_s$, $\alpha_{\rm gw}$, and
$\delta_{\rm gw}$ and on the observed direction $(\alpha,\delta)$:
$\ve{V}_{c/s}=\ve{V}_{c/s}(\alpha,\delta\,;\,h^+_c, h^+_s, h^\times_c,
h^\times_s, \alpha_{\rm gw}, \delta_{\rm gw})$. Omitting the explicit 
dependence on the parameters of gravitational wave and using the notations introduced
in Section~\ref{section-partial-derivatives} one gets
\begin{eqnarray}
\label{V-c}
V^i_c(\alpha,\delta)&=&\delta^i_+\,h^+_c+\delta^i_\times\,h^\times_c\,,
\\
\label{V-s}
V^i_s(\alpha,\delta)&=&\delta^i_+\,h^+_s+\delta^i_\times\,h^\times_s\,.
\end{eqnarray}

The term $\delta t=-{1\over c}\,\ve{p}\cdot\ve{x}_{\rm obs}(t)$ in $\Phi$
appearing in (\ref{delta-u-split}) 
results in a change of the phase under sine and cosine.
For Gaia $\delta t$ is a quasi-periodic function
with a main period of 1 year (due to the motion of Gaia around the
Sun) and an amplitude of maximally 514 seconds.  The frequencies $\nu$
that
are of interest for Gaia are such that the corresponding period 
exceeds 1.5 rotational periods of Gaia (see
Section~\ref{section-frequency-limits} below):
$\nu\le3\times10^{-5}\ {\rm Hz}$. This means that $\delta t$
leads to a quasi-periodic change of the phase with a main period of 1
years and the maximal amplitude of $|2\pi\,\nu\,\delta t|<0.1\,{\rm 
rad}=5.8^\circ$. This effect leads to a change of the gravitational wave signal
of a magnitude of maximally
10\% of the main effect for the highest considered frequencies $\nu$. The effect is 
small enough and can be neglected as a first approximation, which corresponds to considering
a fictitious observer located at the solar system barycentre.
In this way we separate the dependence on time from the angular one:
\begin{eqnarray}
\label{delta-u-split-approximated}
\delta u^i&\approx&V^i_c\,\cos\Phi_0+V^i_s\,\sin\Phi_0\,,
\\
\label{Phi0}
\Phi_0&=&2\pi\,\nu\,t\,.
\end{eqnarray} 
\noindent
This approximation can be used to search for the signal and to
get rough estimates of the parameters of gravitational waves. For the
refinement of the parameters, the full model should be used again (see below).

The first goal of the data processing in this approach is to estimate
$\ve{V}_c$ and $\ve{V}_s$ for reasonably many directions on the sky.
Observational data are residuals of the standard astrometric solution
and can be interpreted as values of $\delta u^i$ disturbed by
observational noise and errors of the astrometric solution itself. If
the frequency $\nu$ is assumed, $\Phi_0(t)$ is known, so that vector
fields $\ve{V}_c$ and $\ve{V}_s$ for a given direction
$(\alpha,\delta)$ can be estimated provided that sufficient number of
observations are available. Eq.~(\ref{delta-u-split-approximated})
taken for one moment of time gives at most 2 equations (see below) for
4 unknown components of $\ve{V}_c$ and $\ve{V}_s$ (recall that
$\delta\ve{u}$, $\ve{V}_c$, and $\ve{V}_s$ are orthogonal to
$\ve{u}$). Therefore, one observation for a given $(\alpha,\delta)$ is
not sufficient to estimate $\ve{V}_c$ and $\ve{V}_s$.

At any given
moment of time an astrometric instrument observes only relatively
small region(s) of the sky. In particular, Gaia has two fields of view
and at any moment of time observes sources within two regions on the
sky of about $0.7^\circ\times0.7^\circ$. Each direction on the sky is
observed many times at different epochs, so that usual
source parameters -- positions, proper motions and parallaxes -- can
be obtained from the data.
In order to estimate $\ve{V}_c$ and $\ve{V}_s$ for a given position
$(\alpha,\delta)$ on the sky one can combine (1) all observations of a
given source or (2) observations of all sources within a given pixel on
the sky. In the first approach, one determines $\ve{V}_c$ and
$\ve{V}_s$ for the position of a given source at some reference epoch
and neglects all variations of the apparent position of that source.
In the second one, $\ve{V}_c$ and
$\ve{V}_s$ are estimated for the centre of a pixel on the sky and the
variation of the gravitational wave signal within that pixel is
neglected. Considering that the gravitational wave signal is a
large-scale pattern slowly changing over the sky (see
Section~\ref{sect-VSH-expansion}), one can hope to reach a good level of
approximation even if relatively large pixels (e.g. of
several degrees) are used. This way to determine $\ve{V}_c$ and $\ve{V}_s$
for given $(\alpha,\delta)$ would work for step-stare astrometry with
two-dimensional observations.

Astrometric scanning instruments like Gaia bring one more
complication: the observations
have strong asymmetry in the accuracies along and across the scanning
direction \citep{LindegrenEtAl2012,PrustiEtAl2016}. Moreover,
the across-scan observations are used to determine the attitude of the satellite and cannot be used
in further fits.
Therefore,  only
along-scan observations should be used and
Eq.~(\ref{delta-u-split-approximated}) should be modified to give only
one equation per observation:
\begin{eqnarray}
\label{delta-u-split-approximated-AL}
\delta_{\rm AL}=\ve{s}\cdot\delta\ve{u}&\approx&\ve{s}\cdot\ve{V}_c\,\cos\Phi_0+\ve{s}\cdot\ve{V}_s\,\sin\Phi_0\,,
\end{eqnarray} 
\noindent
where $\ve{s}=\ve{s}(t)$ defines the scan direction, so that
$\delta_{\rm AL}$ is the along-scan effect of the gravitational wave.
Each position on the sky is observed many times at different moments
of time $t$ and with different scan directions $\ve{s}(t)$ as
prescribed by the observational schedule called ``scanning law''
\citep{PrustiEtAl2016}.  This means that for each moment of time only
a projection of the vector fields on the scanning direction in two
observing directions can be seen. The vector fields $\ve{V}_c$ and
$\ve{V}_s$ should then be restored from a set of projections for
different moments of time.  This again can be done either for each
source separately or for some pixels on the sky.

In case of scanning instruments and in particular for Gaia with its
huge amount of observations (about $10^{12}$ for 5 years of
observations), it is advantageous to compress the residuals of
astrometric solution by computing a weighted mean of the along-scan
residuals from all observations obtained in a given field of view
within some short interval of time.  Within this small interval of
time both the observed direction as well as the scanning direction can
be considered as constant. This effectively defines observational
normal points and significantly decreases the volume of data to be
used e.g. to determine $\ve{V}_c$ and $\ve{V}_s$. For example, instead
of $10^{12}$ observations expected from Gaia within 5 years of nominal
mission, one gets only about $3\times10^{8}$ normal points for time
intervals of 1 s or only $2\times10^7$ normal points for 15 s.  The
duration of the time intervals is a parameters that can be optimized.
Each normal point consists of averaged sky position
$(\overline{\alpha},\overline{\delta})$, averaged reference time
$\overline{t}$, averaged scan direction $\overline{\ve{s}}$ and
averaged along-scan residual $\overline{r}_{\rm AL}$. According to
$(\overline{\alpha},\overline{\delta})$ these normal points can be
then attributed to a sky pixel, so that several data points are used
to estimate $\ve{V}_c$ and $\ve{V}_s$ as described above.

In this way, we compress the data in two steps: (1) producing averaged normal points
over certain time intervals, and (2) computing $\ve{V}_c$ and $\ve{V}_s$ from those
normal points for some pixels on the sky. Note that the first step is independent
of the model of gravitational wave, e.g. independent of the assumed frequency $\nu$ of the
gravitational wave, while the second step should be repeated for each frequency $\nu$
that needs to be checked.

Once determined for an assumed frequency $\nu$, $\ve{V}_c$ and $\ve{V}_s$ can be analyzed
using usual scheme of the VSHs. These vector fields and in particular
their VSH expansions contain all the information needed to estimate
six parameters of the gravitational wave $h^+_c$, $h^+_s$,
$h^\times_c$, $h^\times_s$, $\alpha_{\rm gw}$, $\delta_{\rm gw}$ or
conclude that there is no statistically significant signal for a given
frequency \citep[Section~5.2]{MignardKlioner2012}. To speed up the VSH
analysis it is of great advantage to use an iso-latitude sky
pixelization scheme like HEALPix \citep{GorskyEtAl2005}.

This VSH analysis should be performed for a grid of frequencies $\nu$.
The computations for different frequencies are obviously independent from
each other. The overall algorithm is therefore embarrassingly
parallel and the data are compact enough to allow a quick and efficient
search for gravitational waves in astrometric data. Moreover, one can further
optimize the algorithm in the style of FFT by using an equidistant grid in frequencies
$\nu_k=k\,\Delta\nu$, $k=1,\dots K$, and the standard recurrence formulas
for $\cos\nu_k$ and $\sin\nu_k$ in terms of $\cos\Delta\nu$ and $\sin\Delta\nu$.
In this way each computational node would compute the fits of
$\ve{V}_c$ and $\ve{V}_s$ for all considered frequencies for a certain HEALPix pixel.

Thus the proposed detection algorithm consists in (1) computing
normal points of the along-scan residuals for some 
sufficiently short time intervals, (2) computing averaged values of
the vector fields $\ve{V}_c$ and $\ve{V}_s$ over a HEALPix pixels on
the sky for a grid of frequencies $\nu_k$, and (3) VSH analysis of the
computed vector fields against the analytical model of the astrometric
signal of a gravitational wave. In the case of detection, the
algorithm delivers preliminary estimates of all 7 parameters of
the gravitation wave. These values can be used
for the final optimization of all 7 parameters of the gravitational
wave using a robust least-square fit directly in AGIS or some local
non-linear optimization (e.g. Leverberg-Marquardt \citep{NumRes1992})
in a separate data processing step. At this last stage of the
parameter determination, the approximations of the search algorithm --
ignoring the additional term in the phase of the gravitation wave
depending of the barycentric position of observer, averaging the
along-scan residuals over certain intervals of time, estimating an
averaged values of the vector fields $\ve{V}_c$ and $\ve{V}_s$ for
some sky pixels -- are no longer used.
Even if these assumptions bias the initial estimates of the gravitational wave
parameters, the last stage eliminates those biases.

Obviously, the consequences of all the approximations used in the
algorithm should be carefully investigated. Further details of the
algorithm, its implementation and performance will be published
elsewhere.

In the preprint \citep{2017arXiv170706239M} the authors suggest to use
a Bayesian technique to search for gravitational waves in the
astrometric data. In order to speed up the Bayesian search that is
relatively slow in its nature, the authors suggest to compress the
data before the search by replacing the actual observations with their
unweighted average over certain Voronoi cells on the sky defined
through a set of artificial points. In principle, in spite of obvious
differences, this idea is similar to the use of the HEALPix sky
pixelization to speed up data analysis and VSH expansions which is
one of the standard applications of the HEALPix scheme
\citep{GorskyEtAl2005} and also used in the present paper. Unfortunately, the
work \citep{2017arXiv170706239M} fails to adequately account for
important properties of Gaia observations.  In particular, the authors
consider that (1) observations of all stars are performed
simultaneously, (2) the observations are two-dimensional, (3) the
interaction with the astrometric solution occurs only via proper
motions (namely, the authors ignore parallaxes as well as attitude and
calibration parameters). Finally the model for observational
uncertainties of Gaia data is too simplistic. For all these reasons
the results of \citep{2017arXiv170706239M}, while being interesting
for the selected toy model, are not adequate for the real Gaia
astrometry.

\section{VSH expansion of the astrometric signal of a gravitational wave}
\label{sect-VSH-expansion}

We analyze now the properties of VSH expansions of the vector fields
$\ve{V}_c$ and $\ve{V}_s$ resulting from a plane gravitational wave.
The VSH formalism used below is formulated in \citep{MignardKlioner2012}
and is based on the standard VSH theory exposed e.g. in \citep{GelfandMilnosShapiro1963}.
Each of the two vector fields can be represented as
\begin{equation}\label{Vexpand}
    \ve{V}(\alpha,\delta) = \sum_{l=1}^\infty\,\sum_{m=-l}^{l}\, 
\bigl(t_{lm} \mathbf{T}_{lm} + s_{lm} \mathbf{S}_{lm}\bigr)\,,
\end{equation}
\noindent
where $\mathbf{S}_{lm}$ and $\mathbf{T}_{lm}$ are the spheroidal and
toroidal VSHs, and $t_{lm}$ and $s_{lm}$ are the corresponding
time-independent coefficients that can be estimated from the
data for $\ve{V}$. Since the vector fields $\ve{V}_c$ and $\ve{V}_s$ are real, the
expansion can be simplified as
\begin{eqnarray}\label{Vexpandreal}
    \ve{V}(\alpha,\delta) &=&  \sum_{l=1}^\infty\,\Biggl(
t_{l0} \ve{T}_{l0} + s_{l0} \ve{S}_{l0}
+ 2\sum_{m=1}^{l}\, \left(t^{\Re}_{lm} \ve{T}^{\Re}_{lm} -  t^{\Im}_{lm} \ve{T}^{\Im}_{lm}
+s^{\Re}_{lm} \ve{S}^{\Re}_{lm} -  s^{\Im}_{lm} \ve{S}^{\Im}_{lm}
\right)\Biggr)\,,
\end{eqnarray}
\noindent
where subscripts $\Re$ and $\Im$ denote real and imaginary parts of complex quantities.
In particular, $\ve{T}^{\Re}_{lm}=\Re(\ve{T}_{lm})$, $\ve{T}^{\Re}_{lm}=\Im(\ve{T}_{lm})$,
$\ve{S}^{\Re}_{lm}=\Re(\ve{S}_{lm})$, $\ve{S}^{\Im}_{lm}=\Im(\ve{S}_{lm})$.
The coefficients $t^{\Re}_{lm}$, $t^{\Im}_{lm}$, $s^{\Re}_{lm}$, and $s^{\Im}_{lm}$ are real numbers 
(note that $t^{\Im}_{l0}=s^{\Im}_{l0}=0$) and are defined as
\begin{eqnarray}
\label{t-Re-lm}
t^{\Re}_{lm} &=&  \int_S\,  \ve{V}\cdot \ve{T}^{\Re}_{lm}\,dS\,,\\
\label{t-Im-lm}
t^{\Im}_{lm} &=&  -\int_S\,  \ve{V}\cdot \ve{T}^{\Im}_{lm}\,dS\,,\\
\label{s-Re-lm}
s^{\Re}_{lm} &=&  \int_S\,  \ve{V}\cdot \ve{S}^{\Re}_{lm}\,dS\,,\\
\label{s-Im-lm}
s^{\Im}_{lm} &=&  -\int_S\,  \ve{V}\cdot \ve{S}^{\Im}_{lm}\,dS\,,
\end{eqnarray}
\noindent
where for any function $f$ the integral $\int_Sf\,dS$ is computed over the whole sphere as
$\int_Sf\,dS = \int\limits_0^{2\pi}d\alpha\,\int\limits_{-\pi/2}^{\pi/2}d\delta\,\cos\delta\,f\,$.
Further details can be found e.g. in \citep{MignardKlioner2012,Klioner2012}.

In order to investigate the explicit form of the VSH coefficients in
(\ref{Vexpandreal}) as functions of the parameters of the gravitational
wave, it is advantageous to write
Eq.~(\ref{delta-u-split-approximated}) as
\begin{eqnarray}
\label{delta-u-split-approximated-analysis}
\delta u^i&\approx&\delta^i_+\,\left(h^+_c\,\cos\Phi_0+h^+_s\,\sin\Phi_0\right)
+\delta^i_\times\,\left(h^\times_c\,\cos\Phi_0+h^\times_s\,\sin\Phi_0\right)\,,
\end{eqnarray} 
\noindent
and then compute the VSH expansion (\ref{Vexpandreal}) with
(\ref{t-Re-lm})--(\ref{s-Im-lm}) for two vector fields $\delta^i_+$
and $\delta^i_\times$ that depend only on the position on the sky
$(\alpha,\delta)$ and on the propagation direction $(\alpha_{\rm
  gw},\delta_{\rm gw})$ of the gravitational wave as parameters.
Moreover, the transformation laws of the VSH coefficients
$(t_{lm},s_{lm})$ under spatial rotations given in detail e.g. in
Section~3 of \citep{MignardKlioner2012}, can be used to obtain the
VSH expansion in arbitrary reference system. Therefore,
the theoretical analysis here can consider the simplest case and assume that
$\delta_{\rm gw}=\pi/2$ and $\alpha_{\rm gw}=3\pi/2$. In this case, $\ve{P}$ is
the unit matrix, $\ve{p}=(0,0,1)$ and one gets
\begin{eqnarray}
\delta^i_+=
\begin{pmatrix}
-\cos\alpha\,\cos\delta\,\left(\sin^2\alpha+{1\over 2}\cos2\alpha\,\sin\delta\right)\\[3pt]
\sin\alpha\,\cos\delta\,\left(\cos^2\alpha-{1\over 2}\cos2\alpha\,\sin\delta\right)\\[3pt]
{1\over 2}\,\cos2\alpha\,\cos^2\delta
\end{pmatrix}
={1\over 2}\,\sin2\alpha\,\cos\delta\,e^i_\alpha+{1\over 2}\,\cos2\alpha\,\cos\delta\,e^i_\delta
\,,
\\[3pt]
\delta^i_\times=
\begin{pmatrix}
\sin\alpha\,\cos\delta\,\left(\cos^2\alpha\,(1-\sin\delta)-{1\over 2}\right)\\[3pt]
\cos\alpha\,\cos\delta\,\left(\sin^2\alpha\,(1-\sin\delta)-{1\over 2}\right)\\[3pt]
{1\over 2}\,\sin2\alpha\,\cos^2\delta
\end{pmatrix}
=-{1\over 2}\,\cos2\alpha\,\cos\delta\,e^i_\alpha+{1\over 2}\,\sin2\alpha\,\cos\delta\,e^i_\delta
\,.
\end{eqnarray}
\noindent
where $e_\alpha^i$ and $e_\delta^i$ are vectors of the local triad:
\begin{eqnarray}
\label{ni-alpha-delta}
\ve{u}&=&\begin{pmatrix}\cos\alpha\,\cos\delta\\
\sin\alpha\,\cos\delta\\
\sin\delta
\end{pmatrix}\,,
\\
\label{e_alpha}
\ve{e}_\alpha&=&{1\over\cos\delta}\,{\partial\over\partial\alpha}\ve{u}
=\begin{pmatrix}-\sin\alpha\\ \phantom{-}\cos\alpha\\ 0\end{pmatrix}\,,
\\
\label{e_delta}
\ve{e}_\delta&=&\ve{u}\times\ve{e}_\alpha={\partial\over\partial\delta}\,\ve{u}
=\begin{pmatrix}-\cos\alpha\,\sin\delta\,\\ -\sin\alpha\,\sin\delta\\ \cos\delta\end{pmatrix}\,.
\end{eqnarray}

\begin{figure}[htb]
  \hbox{
    \hskip-1.2cm
    \vbox{
 \includegraphics[width=12.cm,clip=true]{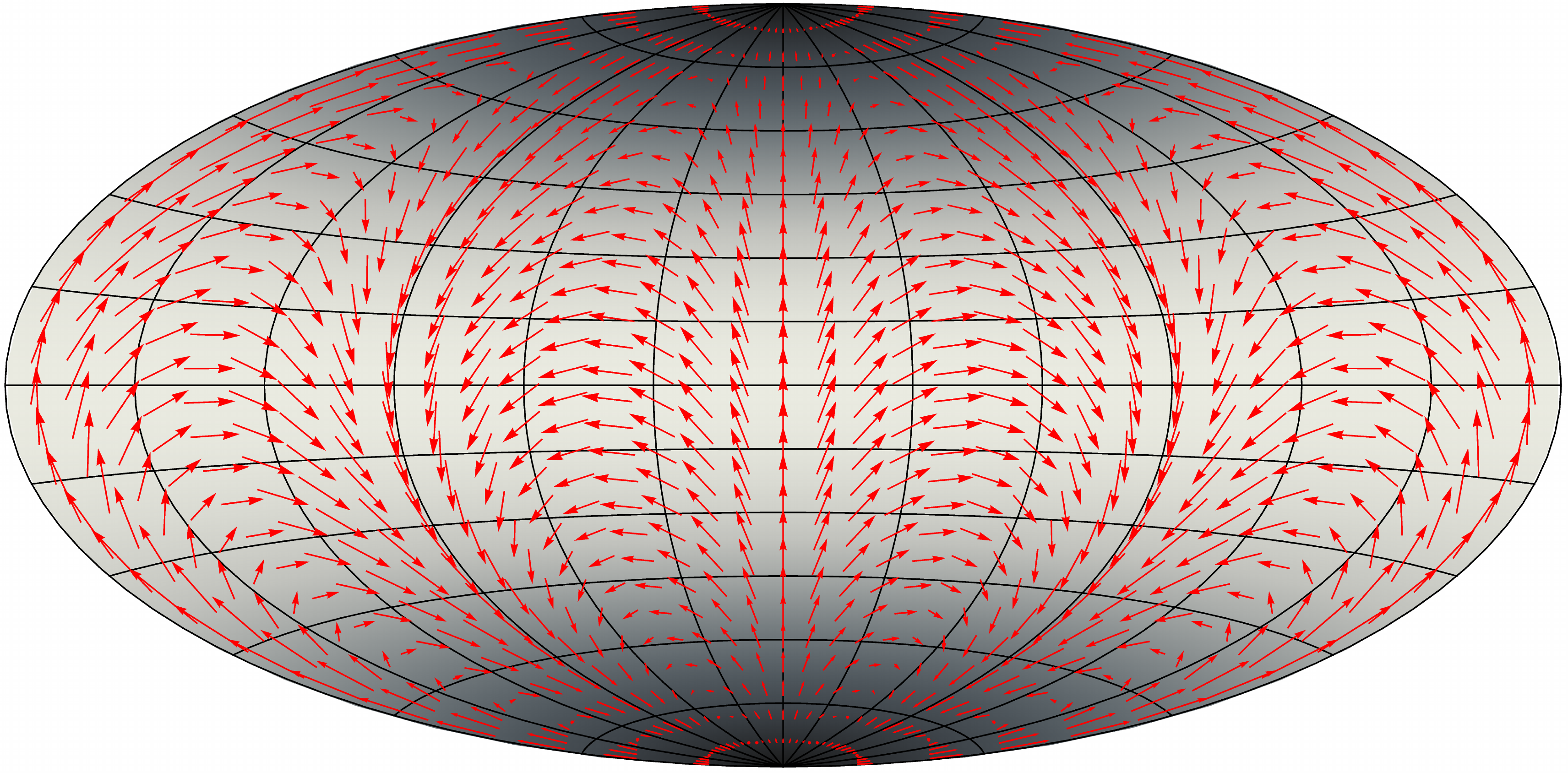}  
 \vskip3mm
 \includegraphics[width=12.cm,clip=true]{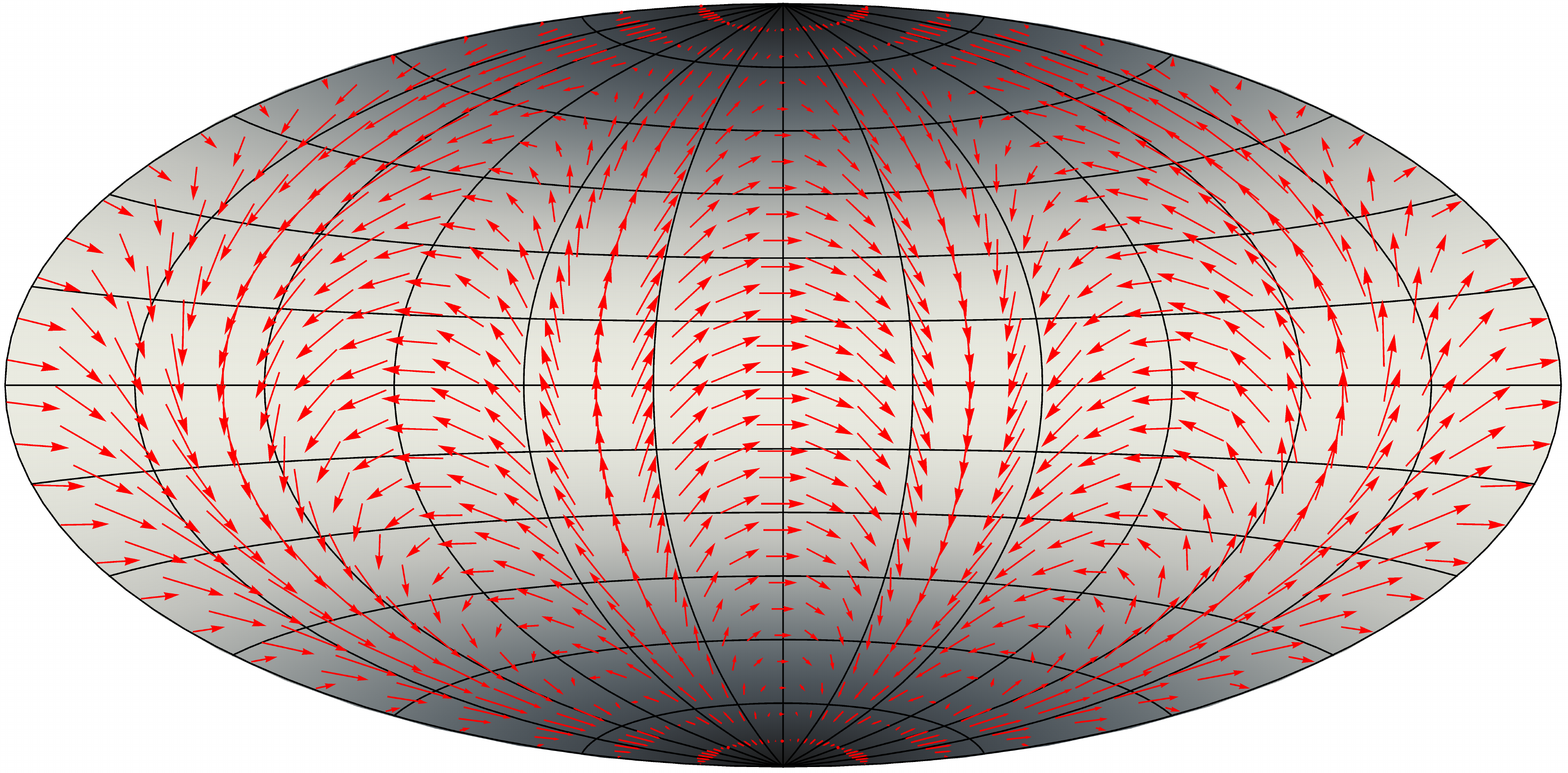}
    }
    \hskip-1cm
   \raisebox{25mm}{\includegraphics[width=1cm,clip=true]{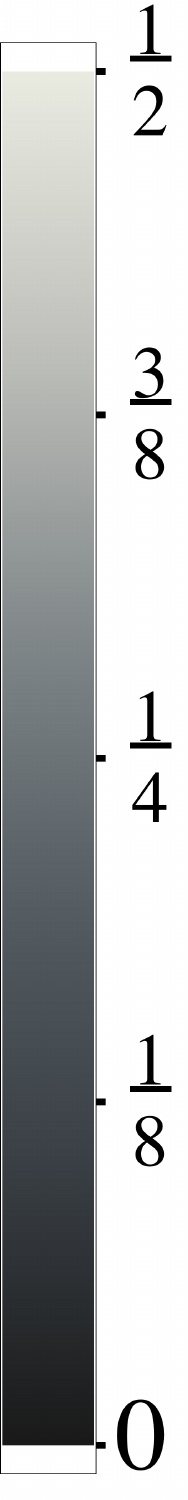}}
    }
\caption{Vector fields $\delta^i_+$ (upper pane) and $\delta^i_\times$
  (lower pane) for a gravitational wave propagating towards the north
  pole $\delta_{\rm gw}=\pi/2$. The maps use an Aitoff projection in equatorial coordinates
  $(\alpha,\delta)$, with origin $\alpha=\delta=0$ at the centre
  and $\alpha$ increasing from right to left. The gray-scale in the background shows the
  magnitude of the vector field (the lighter the bigger), which is equal to
  $\frac{1}{2}\,\cos\delta$ in both cases.
}
       \label{figure-delta-plus-times}
\end{figure}

\begin{figure}[htb]
\centering
\includegraphics[width=12.cm,clip=true]{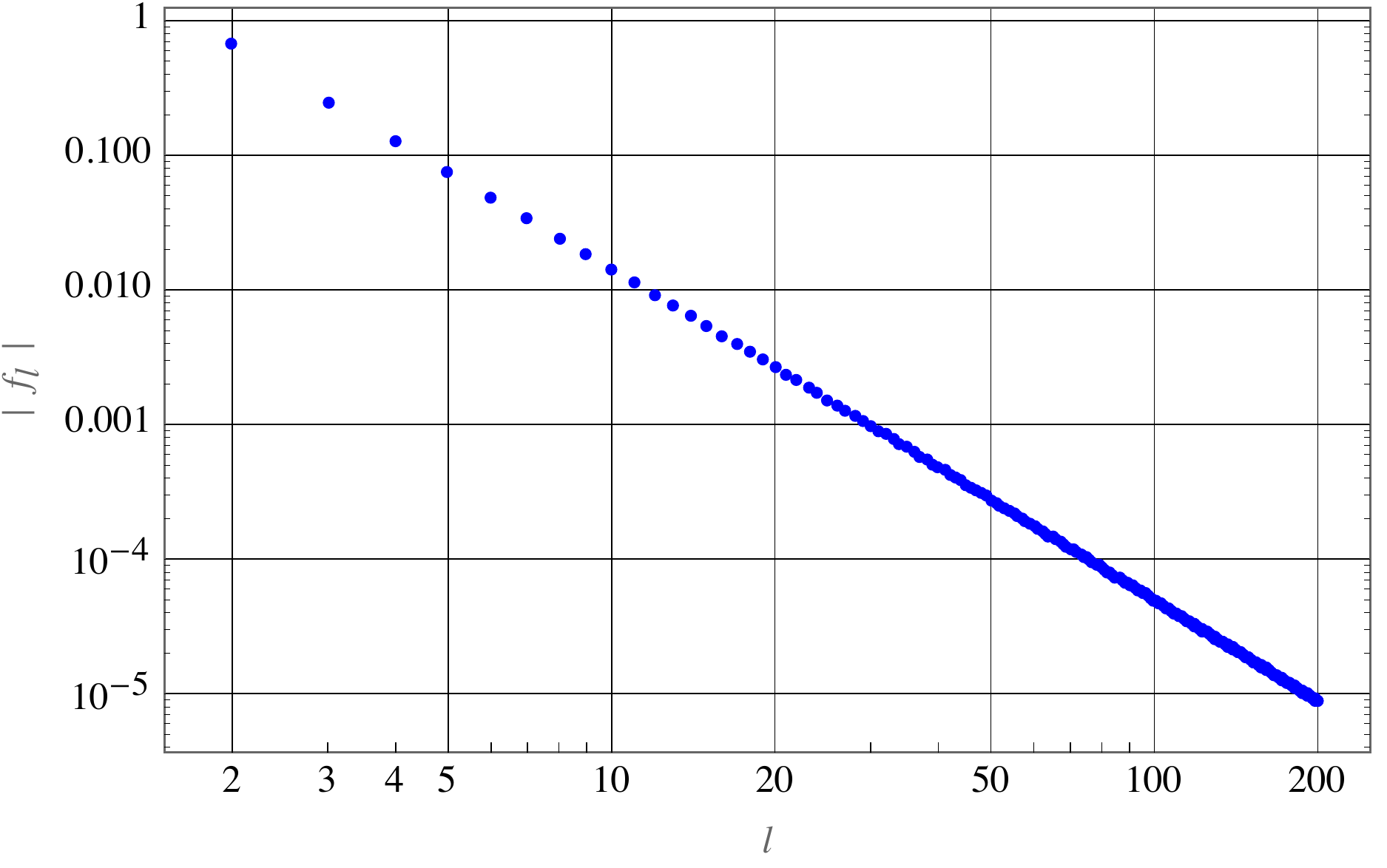}
\caption{The absolute values of the coefficients $f_l$. 
  The values of $f_l$ are related to the relative amplitudes of the astrometric
  signal from a gravitational wave at different VSH orders as given by
  Eqs.~(\ref{veVc-expansion})--(\ref{veVs-expansion}) and
  are more relevant for astrometric detection than the values of $g_l$ shown on Fig.~1 of \citep{BookFlanagan2011}.}
       \label{figure-fl}
\end{figure}

Note that vectors $\delta^i_\times$ and $\delta^i_+$ are orthogonal to
each other and to vector $u^i$ at any point $(\alpha,\delta)$, and
$|\delta^i_\times|=|\delta^i_+|={1\over 2}\,\cos\delta$. These
vectors can be directly used in (\ref{t-Re-lm})--(\ref{s-Im-lm}). Some
partial results can be found in \citep{BookFlanagan2011}. In general,
the non-zero VSH coefficients
in the expansion (\ref{Vexpandreal}) for $\delta^i_\times$ and $\delta^i_+$ read:
\begin{eqnarray}
\label{deltaPlus-t-Re-lm}
&&{\rm for }\ \delta^i_+:\qquad t^{\Im}_{l2}=-s^{\Re}_{l2}=f_l,\ l\ge2\,,\\
\label{deltaTimes-t-Re-lm}
&&{\rm for }\ \delta^i_\times:\qquad t^{\Re}_{l2}=+s^{\Im}_{l2}=f_l,\ l\ge2\,,\\[5pt]
\label{al}
&&f_l =(-1)^l\,{2\over l\,(l+1)}\,\sqrt{(2l+1)\,\pi\over(l-1)\,(l+2)},\ l\ge2\,. 
\end{eqnarray}
\noindent
All other VSH coefficients vanish (this includes also all coefficients with $l=1$).
We note here that according to the definition of
Eqs.~(\ref{deltaPlus-t-Re-lm})--(\ref{deltaTimes-t-Re-lm}),
$f_l$ for $l=1$ is undefined.
Coefficients $f_l$ are depicted on Fig.~\ref{figure-fl}.
The vector fields $\delta^i_\times$ and $\delta^i_+$ are depicted on
Fig \ref{figure-delta-plus-times}. General
deflection pattern due to a plane gravitational wave is a
time-dependent linear combination
(\ref{delta-u-split-approximated-analysis}) of two shown vector
fields considered in a suitable spatial orientation. 
In particular, the VSH expansion of the vector fields 
and 
the rotation-invariant
powers of the toroidal and spheroidal components of the degree $l$ 
\citep[Section~5.2]{MignardKlioner2012}
are equal to each other and read
\begin{eqnarray}
&&\ve{V}_c = 2\sum_{l=2}^\infty\,
f_l \left(h^\times_c\,\left(\ve{T}^{\Re}_{l2} - \ve{S}^{\Im}_{l2}\right)
-h^+_c\,\left(\ve{T}^{\Im}_{l2}+\ve{S}^{\Re}_{l2}\right)
\right)\,,
\label{veVc-expansion}
\\
&&\ve{V}_s = 2\sum_{l=2}^\infty\,
f_l \left(h^\times_s\,\left(\ve{T}^{\Re}_{l2} - \ve{S}^{\Im}_{l2}\right)
-h^+_s\,\left(\ve{T}^{\Im}_{l2}+\ve{S}^{\Re}_{l2}\right)
\right)\,,
\label{veVs-expansion}
\\
&&\left.P_l^t\right|_c=\left.P_l^s\right|_c=2\,f_l^2\left(\left(h^+_c\right)^2+\left(h^\times_c\right)^2)\right)
\,,
\\
&&\left.P_l^t\right|_s=\left.P_l^s\right|_s=2\,f_l^2\left(\left(h^+_s\right)^2+\left(h^\times_s\right)^2)\right)\,.
\end{eqnarray}
\noindent
Note that
\begin{equation}
\sum_{l=2}^\infty f_l^2={\pi\over6}\,
\end{equation}
\noindent 
and the relative power of the toroidal and spheroidal terms (both separately and the sum of them) 
at order $l$ for both vector fields reads
\begin{eqnarray}
\label{gl}
g_l&=&P_l^t/\sum_{l=2}^\infty P_l^t=P_l^s/\sum_{l=2}^\infty P_l^s
=(P_l^t+P_l^s)/\sum_{l=2}^\infty(P_l^t+P_l^s)
\nonumber\\
&=&\left({\pi\over6}\right)^{-1}\,f_l^2={24\,(2l+1)\over(l-1)\,l^2\,(l+1)^2\,(l+2)}\,,\ l\ge2\,.
\end{eqnarray}
\noindent
This gives a simple explicit formula for the coefficients
that \citet{BookFlanagan2011} denoted as $\alpha_l^{EE}$ and $\alpha_l^{BB}$ and 
calculated numerically: $\alpha_l^{EE}=\alpha_l^{BB}=g_l$. This generalizes the results of 
\citep{BookFlanagan2011,PyneEtAl1996}.

Having the observational data for $\ve{V}_c$ and $\ve{V}_s$ for
certain locations on the sky (see Section~\ref{section-vhs}), one can
fit the coefficients $t_{lm}$ and $s_{lm}$ using a sort of least
squares estimator. Many details of this procedure is given in
e.g. Section~5 of \citep{MignardKlioner2012}. In this way for a given
frequency $\nu$ one gets two sets of the coefficients
$(t^c_{lm},s^c_{lm})$ and $(t^s_{lm},s^s_{lm})$ for $\ve{V}_c$ and
$\ve{V}_s$, respectively. These sets of coefficients can be used to
decide if a signal from a gravitational wave is detected in the data
(see again Section~5 of \citep{MignardKlioner2012}). In addition to
the standard statistical criteria, the symmetries of the signal and in
particular the fact that the power is equal in the toroidal and
spheroidal harmonics at each order as well as the decrease of the
power with $l$ given by (\ref{gl}) can be used to distinguish the
signal due to gravitational waves from other kinds of signals. If a
signal is detected for some frequency $\nu$ the VSH coefficients can
be converted back to the 6 parameters $h^+_c$, $h^+_s$, $h^\times_c$
and $h^\times_s$, $\alpha_{\rm gw}$, and $\delta_{\rm gw}$ using the
model above and the transformation laws
of the coefficients under rotations as described e.g. in
\citep[Section~3]{MignardKlioner2012}. To increase the signal-to-noise
ratio when computing $(\alpha_{\rm gw},\delta_{\rm gw})$ one can also
analyze the formal sum $\ve{V}_c+\ve{V}_s$. As discussed in Section~\ref{section-vhs}
these estimated values of the gravitational wave parameters should be used
for a final fit of those parameters directly in the astrometric solution.

\section{Possible frequency limits and sensitivity of Gaia astrometry}
\label{section-frequency-limits}

The frequency $\nu$ of the signal that can be potentially detected by
Gaia is not arbitrary.  The upper limit for $\nu$ comes from the fact
that the Gaia instrument should be calibrated from the same data.
Although the details of the calibration are not fully known it is
likely that the calibration will attempt to eliminate all periodic
signals in the data with periods smaller than 1.5--2 rotational
periods. Considering the best case one concludes that
$\nu\le3\times10^{-5}$~Hz. The lower limit for $\nu$ is related
to the fact that the slow variations of position are well represented by
proper motions -- the first regime discussed in Section~\ref{section-two-regimes}.
In the second regime, which we consider now, the period of gravitational waves
should be (much) smaller than the time span of observations. Again considering the
the best case and the planned mission duration of 5~yr one gets $\nu\ge6.4\times10^{-9}$~Hz.
Finally, one gets
\begin{equation}
6.4\times10^{-9}\ {\rm Hz}\le\nu\le3\times10^{-5}\,{\rm Hz}\,.
\end{equation}
We note here that the lower limit for the frequency can be better (lower) in reality since the
regimes discussed in Section~\ref{section-two-regimes} don't have strict
boundaries. It is clear that if the period of the gravitational wave is exactly
equal to the time span of observations the astrometric solution is unable to adsorb
the corresponding astrometric signal by proper motions which are linear in time.
This is the third regime mentioned in Section~\ref{section-two-regimes}.
Therefore, one can expect that the lower limit for the frequency can be lowered to about
$3\times10^{-9}$~Hz.

\begin{figure}[htb]
\centering
\includegraphics[width=14.cm,clip=true]{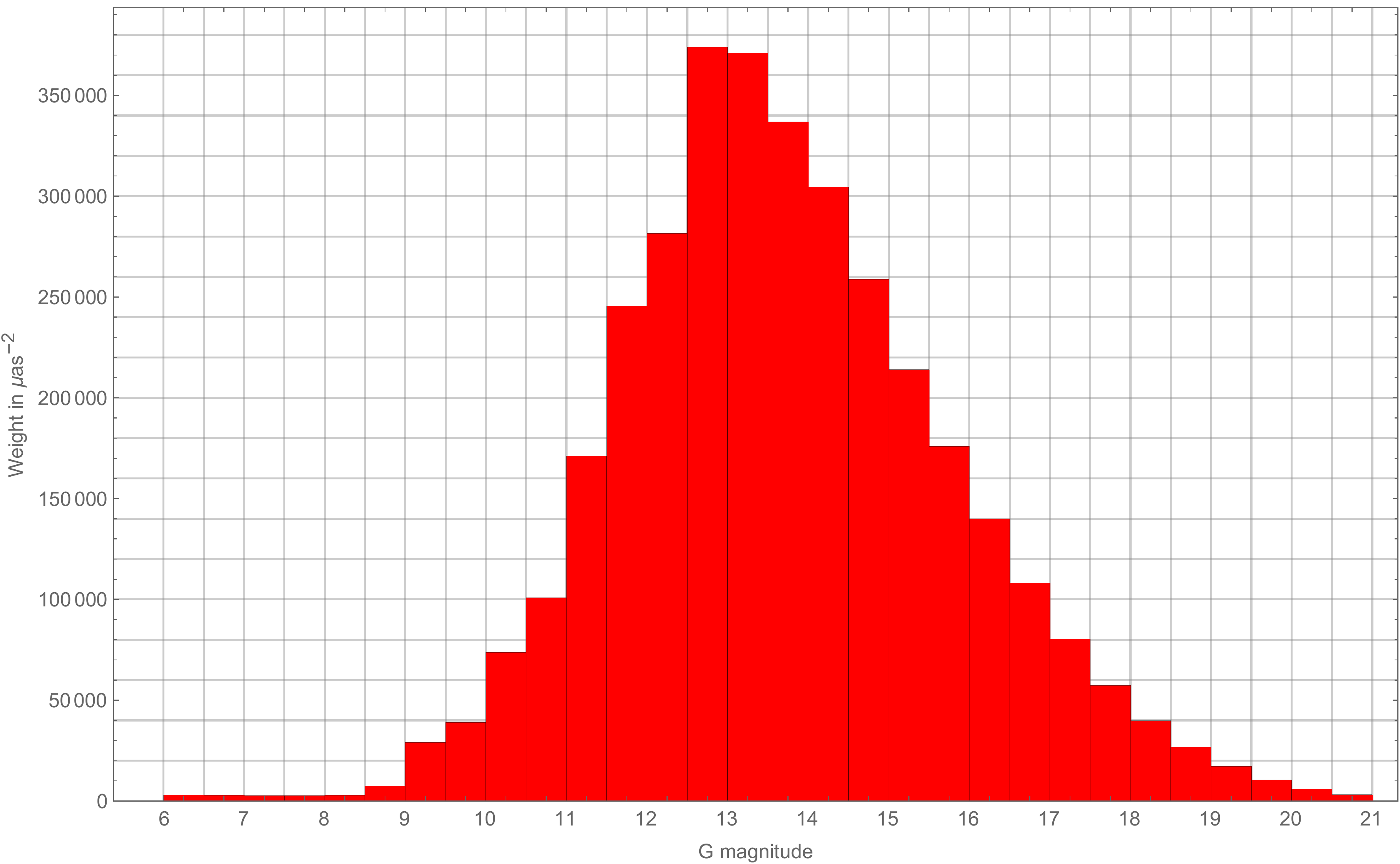}
\caption{The distribution of the statistical weight of Gaia observations as function of Gaia magnitude $G$. Each bin shows the statistical weight of
  observations of sources in the corresponding interval of G. Here the post-launch estimates of
  the errors of Gaia observations were used. A version of this figure for the pre-launch error estimates
  and for only 10\% of sources can be seen as Fig.~2 of \citep{2010IAUS..261..315H}.
}
\label{figure-StatisticalWeights}
\end{figure}

Another goal of this Section is to discuss the sensitivity that can
be expected from Gaia data. The expected astrometric accuracy of Gaia
observations can be found at
\url{http://www.cosmos.esa.int/web/gaia/science-performance#astrometric%20performance}.
This accuracy can be converted to the expected uncertainties of a single observation of Gaia
as function of the Gaia magnitude $G$ as defined \citep{2010A&A...523A..48J}. Then a model  
of the Universe \citep{2012A&A...543A.100R} can be used to compute the number of
sources expected in each small interval in $G$. Combining the uncertainties of observations
with the star counts one can calculate that the total statistical
weight of all Gaia observations for stars up to magnitude $G=20$ reads
\citep{KlionerSteidelmueller2012,Klioner2015}
\begin{equation}
W_{\rm full}=3.4\times10^6\,{\muas}^{-2}\,.
\end{equation}
\noindent
Fig.~\ref{figure-StatisticalWeights} shows the distribution of the
statistical weights of observations in certain intervals in $G$. One
can see that sources with $12\le G\le15$ are the most important for
the determination. As with other global parameters in Gaia (e.g. the
PPN $\gamma$) one can come to the idea to use only bright stars with,
say, $G\le16$. This would considerably reduce the amount of
observations (one can expect about $10^8$ such sources) while almost
retaining the final accuracy of determination: $W_{\rm full}^{G\le
  16}=3.0\times 10^6 \,{\muas}^{-2}$.  However, such a selection of
source is dangerous in the presence of calibration errors that are
often strongly depend on the magnitude and related parameters. The
idea to compute normal points presented above makes such a selection
unnecessary.

We are interested in sensitivity of Gaia data to
the overall amplitude of the gravitational wave
$h=\sqrt{\left(h^+_c\right)^2+\left(h^+_s\right)^2+\left(h^\times_c\right)^2+\left(h^\times_s\right)^2}$.
For any scalar parameter $h$ to be fitted to the Gaia data, the maximal possible accuracy 
of its determination is $\left(W_{\rm full}\right)^{-1/2}$. This lower
estimate holds if the partial derivatives
with respect to the parameter are equal to unity for all observations. Therefore,
\begin{equation}
\sigma_h\ge \left(W_{\rm full}\right)^{-1/2}=5.4\times10^{-4}\,\muas=2.6\times10^{-15}\,.
\end{equation}
\noindent
Obviously, the actual sensitivity will be lower because of various correlations and systematic errors. One can write
\begin{equation}  
\sigma_h\approx 2.6\times10^{-15}\,Q\,,
\end{equation}
where $Q$ is a numerical factor depending on the details of Gaia
observations.  As a plain guess it is reasonable to assume that
$Q\sim10-1000$. This factor reflects the fact that the same
observation are used to fit the source, attitude and calibration
parameters (see Section~\ref{section-Introduction}) as well as some
systematic errors.  Note, however, that this should not be interpreted
as a claim of real sensitivity of Gaia.  This only gives an estimate
of the best possible sensitivity.

The actual sensitivity curve (including the actual frequency limits) should
be determined by detailed end-to-end numerical simulations involving in
particular the interaction between the astrometric signal of a
gravitational wave and the standard astrometric solution. The results
of these simulations will be the subject of a separate publication.

\section{Sources of gravitational waves for space astrometry}
\label{section-sources}

It is clear that the most promising sources of gravitational waves for
astrometric detection are supermassive binary black holes in the
centres of galaxies. Recently, such systems were often discussed in
the literature (see, e.g.,
\citep{YonemaruEtAl2016,2015Natur.518...74G,2015MNRAS.453.1562G,2017AAS...22930702M}).  It is
believed that binary supermassive black holes are a relatively common
product of interaction and merging of galaxies in the typical course
of their evolution. This sort of objects can give
gravitational waves with both frequencies and amplitudes potentially
within the reach of space astrometry. Moreover, the gravitational waves
from those objects can often be considered to have virtually constant
frequency and amplitude during the whole period of observations of
several years. A binary system with a chirp mass ${\cal M}$ on a 
circular orbit with the orbital period $P$ emits gravitational waves
of the period $P_{\rm gw}=P/2$ and strain \citep{JaranowskiKrolak2009,Buonanno2007}
\begin{eqnarray}
\label{h-GW}
h&=&{4\,\pi^{2/3}\over c^4}\,{\left(G{\cal M}\right)}^{5/3}\,P_{\rm gw}^{-2/3}\,r^{-1}
\nonumber\\
&=&1.19\times10^{-14}\,
\left({{\cal M}\over 10^9\,M_\odot}\right)^{5/3}\,
\left({P_{\rm gw}\over 1\ {\rm yr}}\right)^{-2/3}\,
\left({r\over 100\ {\rm Mpc}}\right)^{-1}\,,
\end{eqnarray}
\noindent
where $r$ is the (luminosity) distance to the source, $G$ is the Newtonian
constant of gravitation, and $M_\odot$ is the Solar mass. This
equation gives the strain for both
polarizations in the direction perpendicular to the orbital plane.
Two polarizations have different dependence on the
inclination of the orbit \citep{JaranowskiKrolak2009}. 
For eccentric orbits the strain is moderately increased approximately as $h\propto(1-e^2)^{-1}$. However, 
eccentricity of the orbit is not expected to play a big role since 
eccentricity decreases during the evolution and one should generally
expect small eccentricities. The above estimate is derived in the
linear approximation of general relativity, which means that it is valid when
\begin{equation} 
P_{\rm gw}\gg 10^{\rm h}\,\left({{\cal M}\over 10^9\,M_\odot}\right)\,.
\end{equation}
\noindent
This is the condition that the semi-major axis of the orbit is much larger than
the sum of the Schwarzschild radii of the components. It is well known
that such a massive binary system loses energy due to gravitational
radiation, so that its orbital period decreases (inspiralling orbit)
and the frequency of gravitational wave $\nu_{\rm gw}=1/P_{\rm gw}$
increases.  The derivative $\dot \nu_{\rm gw}$ can be computed from the
energy balance between the emitted gravitational waves and the orbital
motion \citep{JaranowskiKrolak2009,Buonanno2007}:
\begin{eqnarray}
\label{f-GW}
\dot \nu_{\rm gw}={96\over 5}\pi^{8/3}\,{\left({G{\cal M}\over c^3}\right)}^{5/3}\,\nu_{\rm gw}^{11/3}
=5.83\times 10^{-12}\ {\rm Hz/yr}\,
\left({{\cal M}\over 10^9\,M_\odot}\right)^{5/3}\,
\left({P_{\rm gw}\over 1\ {\rm yr}}\right)^{-11/3}\,.
\end{eqnarray}
\noindent
Integrating this equation one gets that the time to coalescence
$\tau=t_{\rm coal}-t_{\rm obs}$ (in this approximation $\nu_{\rm gw}$ goes to infinity at the
coalescence) reads \citep{JaranowskiKrolak2009,Buonanno2007}:
%
%
\begin{eqnarray}
\label{tau-GW}
\tau&=&{3\over 8}\,{\nu_{\rm gw}\over \dot \nu_{\rm gw}}={5\over 256}\,\pi^{-8/3}\,{\left({G{\cal M}\over c^3}\right)}^{-5/3}\,P_{\rm gw}^{8/3}
=2039\ {\rm yr}\,
\left({{\cal M}\over 10^9\,M_\odot}\right)^{-5/3}\,
\left({P_{\rm gw}\over 1\ {\rm yr}}\right)^{8/3}\,.
\end{eqnarray}
\noindent
Here $\nu_{\rm gw}$, $\dot \nu_{\rm gw}$ and $P_{\rm gw}$ are evaluated at the moment of observation
$t_{\rm obs}$. Eqs.~(\ref{h-GW}) and (\ref{tau-GW}) give a useful insight of what
sort of binary systems can be within the reach of space
astrometry: the strain (\ref{h-GW}) should be large (say, $\gtrsim10^{-13}$ for Gaia)
and the time to coalescence should be large
enough to guarantee almost constant frequency of gravitational wave
during the whole period of observations (of 5--10 years for Gaia). 

It is clear that the known candidates for binary supermassive black
holes are rather speculative.  Nevertheless, it seems to be useful to
give estimates of the expected strain of the gravitational waves from
those sources. Substituting the corresponding parameters of the
candidates
\citep{2016ApJ...819L..37V,2015Natur.518...74G,YonemaruEtAl2016} into
the formulas above one gets strains of about $h\sim2\times 10^{-16}$
with a period of 6 yr for OJ287 assuming the chirp mass of
$8\times10^8\,M_\odot$, $h\lesssim5\times10^{-16}$ with a period of
2.6 yr for PG\,1302--102, and $h<1.3\times 10^{-12}\,\left(P/\,1\,{\rm
  yr}\right)^{-2/3}$ for M87 assuming equal mass components. Although
we cannot identify promising sources of gravitational waves for Gaia
astrometry now, it is important to note that $h$ is proportional to
${\cal M}^{5/3}/r$ so that moderate increase in the chirp mass can
compensate greater distances. Currently one suspects supermassive
black holes with masses $>10^{10}\,M_\odot$ in a number of
galaxies. Some of them may turn out to be binary systems and represent
sources of gravitational waves for high-accuracy astrometry.

\section{Concluding remarks}
\label{section-conclusion}

In this report we summarized the model for astrometric effects of a
plane gravitational wave with constant frequency. The model and the
most important partial derivatives are given by
Eqs.~(\ref{hij-+})--(\ref{rotation-x}) and
(\ref{Phi})--(\ref{partial-ui-partialOmega}).

The search algorithm based on the data normal points and
VSH analysis described in Sections~\ref{section-vhs} and
\ref{sect-VSH-expansion} is very promising to reduce the
computational complexity of the search for gravitational waves in the
observational data especially in combination with the HEALPix
pixelization \citep{GorskyEtAl2005}. 

In Section~\ref{section-frequency-limits} we gave estimates for the
frequency range in which a Gaia-like instrument can be used to detect
gravitational waves as periodic deflection signals. We also gave an
estimate for the best-case sensitivity of Gaia astrometry. An overview
of the main characteristics of gravitational waves from the binary
supermassive black holes, which obviously represent the most promising
astrophysical sources for space astrometry, is given in Section~\ref{section-sources}.

The simplest version of the gravitational wave model discussed above
is to assume that the frequency $\nu$ is constant. In principle, it is
straightforward to accommodate the search algorithm to the case when
the frequency is a given function of time as e.g. in Eq.~(\ref{f-GW}).
It is sufficient to use this function in (\ref{Phi0}) when computing
vector fields $\ve{V}_c$ and $\ve{V}_s$ using
(\ref{delta-u-split-approximated}) or
(\ref{delta-u-split-approximated-AL}). Obviously one should also
accommodate the time-dependence of the strain parameters: $\ve{V}_c$
and $\ve{V}_s$ are no longer time independent in this case. Since
astrometry is most sensitive to gravitational waves of almost constant
frequency and strain, the time dependence of parameters can be
sufficiently approximated by a linear functions of time. This
generalization is possible, however would increase the number of
parameters to be fitted.

Another important aspect is the situation when several gravitational
waves of comparable amplitudes from different sources are
superimposed. In principle, if a number of strong signals have
different frequencies (which is physically almost guaranteed) no
modification of the algorithm is needed: the signals will be found one
by one.  On the other hand, in the highly unlikely case of two
gravitational waves with equal frequencies coming from different
sources it is difficult to separate them since the sum of two
different quadrupole signals on the sky is equivalent to another one
quadrupole signal with certain parameters. It is, however, doubtful
that this regime is of any practical interest, except for the case of
stochastic background of gravitational waves. The latter case is
beyond the scope of this work.

The search and fit algorithms sketched in Sections~\ref{section-vhs}
and \ref{sect-VSH-expansion} are being further developed and implemented to work with
the real Gaia data in the framework of Gaia Data Processing and
Analysis Consortium (Gaia DPAC). Further details will be published
elsewhere.

\begin{acknowledgements}
I am grateful to Robin Geyer, Uwe Lammers, Alex Bombrun, Lennart
Lindegren, Michael Perryman, and Hagen Steidelm\"uller for numerous
fruitful discussions and continuing interest in the subject. Various
tools and software products produced by the Gaia DPAC were used in
this work and are gratefully acknowledged. I thank the anonymous
referees for their comments and suggestions that helped to improve the
paper. This work was partially supported by the BMWi grant
50\,QG\,1402 awarded by the Deutsche Zentrum f\"ur Luft- und Raumfahrt
e.V. (DLR) as well as by the ESA under Contract
No. 4000115263/15/NL/IB.
\end{acknowledgements}



\begin{thebibliography}{999}

\bibitem[Blanchet et al.(2001)]{2001LNP...562..141B}
  Blanchet, L., Kopeikin, S., \& Sch{\"a}fer, G.\ 2001,
  in: Gyros, Clocks, Interferometers: Testing Relativistic Gravity in Space,
Springer, Berlin, p.141
  
\bibitem[Book \& Flanagan(2011)]{BookFlanagan2011}
Book, L.G., Flanagan, \'E.\'E. 2011,
Phys.Rev.D 83, 024024

\bibitem[Braginsky {\it et al.}(1990)]{BraginskyEtAl1990} 
Braginsky, V.B., Kardashev, N.S., Polnarev, A.G., Novikov, I.D. 1990,
Nuovo Cimento Soc. Ital. Fis. 105B, 1141

\bibitem[Buonanno(2007)]{Buonanno2007}
Buonanno, A., 2007, Gravitational waves, 
available from arXiv:0709.4682, \url{https://arxiv.org/abs/0709.4682}

\bibitem[Butkevich et al.(2017)]{2017A&A...603A..45B} 
Butkevich, A.~G., Klioner, S.~A., Lindegren, L., Hobbs, D., \& van Leeuwen, F.\ 2017, A\&A, 603, A45 

\bibitem[Damour \& Esposito-Far{\`e}se(1998)]{1998PhRvD..58d4003D}
Damour, T., \& Esposito-Far{\`e}se, G.\ 1998, Phys. Rev. D, 58, 044003 

\bibitem[ESA(2000)]{ESA2000}
ESA 2000, GAIA: Composition, Formation and Evolution of
the Galaxy, Technical Report ESA-SCI(2000)4, available at
\url{http://www.rssd.esa.int/doc_fetch.php?id=359232}

\bibitem[Fabricius et al.(2016)]{2016A&A...595A...3F}
  Fabricius, C., Bastian, U., Portell, J., et al.\ 2016, \aap, 595, A3 

\bibitem[Gaia Collaboration, Prusti\ {\it et al.}(2016)]{PrustiEtAl2016}
Gaia Collaboration, Prusti, T. {\it et al.}, \aap,  \textbf{595}, A1 (2016)

\bibitem[Gelfand et al(1963)]{GelfandMilnosShapiro1963}
  Gelfand, I. M., Milnos, R. A., Shapiro, Z.Ya. 1963, Representation of the
Rotation and Lorentz groups (Oxford: Pergamon)

\bibitem[Geyer(2014)]{Geyer2014}
  Geyer, R. 2014, Investigation of Algorithms of Highly Nonlinear Model Fitting on Big Datasets,
  Master Thesis,
  Center for Information Services and High Performance Computing, Technische Universit\"at Dresden
  
\bibitem[Graham et al.(2015a)]{2015Natur.518...74G} 
Graham, M.~J., Djorgovski, S.~G., Stern, D., et al.\ 2015a,
Nature, 518, 74 

\bibitem[Graham et al.(2015b)]{2015MNRAS.453.1562G} 
Graham, M.~J., Djorgovski, S.~G., Stern, D., et al.\ 2015b, 
MNRAS, 453, 1562

\bibitem[G\'orsky {\it et al.}(2005)]{GorskyEtAl2005}
G\'orski, K.M., Hivon, E., Banday, A.J., Wandelt, B.D., Hansen, F.K.,
Reinecke, M., Bartelman, M. 2005, 
Astrophys.J., 622, 759

\bibitem[Gwinn {\it et al.}(1997)]{GwinnEtAl1997}
Gwinn, C.R., Eubanks, T.M., Pyne, T., Birkinshaw, M., Matsakis, D.N. 1997, 
Astroph.J., 485, 87--91

\bibitem[Hobbs et al.(2016)]{2016arXiv160907325H}
Hobbs, D., H{\o}g, E., Mora, A., et al.\ 2016, arXiv:1609.07325 

\bibitem[Hobbs et al.(2010)]{2010IAUS..261..315H}
  Hobbs, D., Holl, B., Lindegren, L., et al.\ 2010, Relativity in Fundamental Astronomy: Dynamics, Reference Frames, and Data Analysis, 261, 315 

\bibitem[Jaranowski \& Kr\'olak (2009)]{JaranowskiKrolak2009}
Jaranowski, P., Kr\'olak, A. 2009,
Analysis of Gravitational-Wave Data,
Cambridge: Cambridge University Press

\bibitem[Jordi et al.(2010)]{2010A&A...523A..48J}
  Jordi, C., Gebran, M., Carrasco, J.~M., et al.\ 2010, A\&A, 523, A48

\bibitem[Klioner(2003)]{Klioner2003}
  Klioner, S.A. 2003,
Astron.J., 125, 1580

\bibitem[Klioner(2004)]{2004PhRvD..69l4001K} 
Klioner, S.~A.\ 2004, Phys.Rev.D, 69, 124001

\bibitem[Klioner(2004)]{Klioner2007} 
Klioner, S.A. 2007, 
in: Lasers, Clocks and Drag-Free: Exploration of Relativistic Gravity in Space, H. Dittus, C. Lämmerzahl, S. G. Turyshev (eds.),
Astrophysics and Space Science Library 349, Springer, Berlin, p.399

\bibitem[Klioner(2012)]{Klioner2012}
Klioner, S.A. 2012, Representation of corrections to source parameters by scalar and vector spherical harmonics, 
GAIA-CA-TN-LO-SK-016, available from the Gaia document archive \url{http://www.rssd.esa.int/llink/livelink}

\bibitem[Klioner(2013)]{Klioner2013}
Klioner, S.A. 2013, Gaia observations and gravitational waves,
GAIA-CA-TN-LO-SK-014, available from the Gaia document archive \url{http://www.rssd.esa.int/llink/livelink}

\bibitem[Klioner(2014)]{Klioner2014}
Klioner, S.A. 2014, Velocity error and effective Basic Angle Calibration (VBAC): basic principles and possible applications, 
GAIA-C3-TN-LO-SK-020, available from the Gaia document archive \url{http://www.rssd.esa.int/llink/livelink} 

\bibitem[Klioner(2015)]{Klioner2015}
Klioner, S.A. 2015, 
in: The Milky Way Unravelled by Gaia: GREAT Science from the Gaia Data
Releases, N.A.Walton, F. Figueras, L. Balaguer-Núñez, C. Soubiran
(eds.), EAS Publication Series, 67--68 (2014) 49, EDP Sciences, Les
Ulis

\bibitem[Klioner \& Steidelm\"uller(2012)]{KlionerSteidelmueller2012}
Klioner, S.A., Steidelm\"uller, H. 2012,
First Results of the Generic Global Update,
available from \url{https://gaia.esac.esa.int/dpacsvn/DPAC/meetings/CU3/AGIS/18-Toulouse-Nov-12/AGIS18-AK&HST-FirstResultsGGU.pdf}

\bibitem[Kopeikin et al.(1999)]{1999PhRvD..59h4023K}
  Kopeikin, S.~M., Sch{\"a}fer, G., Gwinn, C.~R., \& Eubanks, T.~M.\ 1999, Phys. Rev. D, 59, 084023 

\bibitem[Lindegren et al.(2012)]{LindegrenEtAl2012}
Lindegren, L, Lammers, U., Hobbs, D., O'Mullane, W., Bastian, U., Hern\'andez, J. 2012,
A\&A, 538, A78

\bibitem[Lindegren et al.(2016)]{2016A&A...595A...4L}
Lindegren, L., Lammers, U., Bastian, U., et al.\ 2016, A\&A, 595, A4 

\bibitem[Makarov(2010)]{2010IAUS..261..345M}
  Makarov, V.~V.\ 2010, in:
  Relativity in Fundamental Astronomy: Dynamics, Reference Frames, and Data Analysis,
Cambridge: Cambridge University Press, p.345 

\bibitem[Malbet {\it el al.}(2012)]{MalbetEtAl2012}
Malbet, F., L\'eger, A., Shao, M. {\it et al}, 2012,
Experimental Astronomy, 34, 385

\bibitem[Merritt(2017)]{2017AAS...22930702M} 
Merritt, D.\ 2017, 
American Astronomical Society Meeting Abstracts, 229, 307.02

\bibitem[Michalik et al.(2014)]{2014A&A...571A..85M} 
Michalik, D., Lindegren, L., Hobbs, D., \& Lammers, U.\ 2014, A\&A, 571, A85 

\bibitem[MignardKlioner(2010)]{MignardKlioner2010}
Mignard, F., Klioner, S.A. 2010,
in: Relativity in Fundamental Astronomy: Dynamics, Reference Frames, and Data Analysis,
Cambridge: Cambridge University Press, p.306

\bibitem[Mignard \& Klioner(2012)]{MignardKlioner2012}
Mignard, F., Klioner, S.A. 2012, 
A\&A, 547, A59

\bibitem[Moore et al.(2017)]{2017arXiv170706239M}
  Moore, C.~J., Mihaylov, D., Lasenby, A., \& Gilmore, G.\ 2017, arXiv:1707.06239 

\bibitem[Press {\it et al.}(1992)]{NumRes1992}
Press, W.H., Teukolsky, S.A., Vetterling, W.T., Flannery, B.P.\ 1992,
Numerical Recipes (2nd ed.), Cambridge: Cambridge University Press

\bibitem[Pyne {\it et al.}(1996)]{PyneEtAl1996}
Pyne, T., Gwinn, C.R., Birkinshaw, M., Eubanks, T.M., Matsakis, D.N. 1996, 
Astroph.J., 465, 566

\bibitem[Robin et al.(2012)]{2012A&A...543A.100R}
  Robin, A.~C., Luri, X., Reyl{\'e}, C., et al.\ 2012, A\&A, 543, A100 

\bibitem[Schutz(2009)]{Schutz2009}
Schutz, B. 2009,
A First Course in General Relativity,
Cambridge: Cambridge University Press

\bibitem[Schutz(2010)]{Schutz2010}
Schutz, B. 2010,
2010, Relativity in Fundamental Astronomy: Dynamics, Reference Frames, and Data Analysis,
Cambridge: Cambridge University Press, p.234

\bibitem[Soffel et al.(2003)]{SoffelEtAl2003}
Soffel, M., Klioner, S.~A., Petit, G., et al.\ 2003, Astron.J., 126, 2687 

\bibitem[The Theia Collaboration et al.(2017)]{2017arXiv170701348T}
The Theia Collaboration, Boehm, C., Krone-Martins, A., et al.\ 2017, arXiv:1707.01348

\bibitem[Titov \& Lambert(2013)]{TitovLambert2013}
Titov, O., Lambert, S. 2013, 
A\&A, 559, A95

\bibitem[Titov, Lambert \& Gontier(2010)]{TitovLambertGontier2010}
Titov, O., Lambert, S. Gontier, A.-M. 2010,
A\&A, 529, A91

\bibitem[Valtonen et al.(2016)]{2016ApJ...819L..37V} 
Valtonen, M.~J., Zola, S., Ciprini, S., et al.\ 2016, Astrophys.~J.~Lett., 819, L37 

\bibitem[Yonemaru {\it et.al.}(2016)]{YonemaruEtAl2016}
Yonemaru, N., Kumamoto, H., Kuroyanagi, S., Takahashi, K., Silk, J. 2016,
Publications of the Astronomical Society of Japan,
68, 106, DOI: 10.1093/pasj/psw100

\end{thebibliography}
\end{document}